\documentclass[12pt]{article}

\usepackage{amsfonts}
\usepackage{amssymb}
\usepackage{amstext}
\usepackage{amsmath,bm}
\usepackage{amsthm}

\usepackage{caption}
\usepackage{graphicx}

\begin{document}

\def\be{\begin{equation}}
\def\ee{\end{equation}}
\def\ba{\begin{eqnarray}}
\def\ea{\end{eqnarray}}
\def\lb{\label}
\def\nn{\nonumber}

\def\a{\mathbf{a}}
\def\b{\mathbf{b}}
\def\O{\Omega}
\def\L{\mathcal{L}}
\def\r{\bar{\rho}}

\title
{Cosmological aspects of a unified dark 
energy and dust dark matter model}
\author{Denitsa Staicova \and Michail Stoilov\\
{\it  Institute for Nuclear Research and Nuclear Energy}\\Bulgarian Academy of Sciences
}

\date{}

\maketitle

\begin{abstract}
Recently, a model of modified gravity plus single scalar field model was 
proposed, in which the scalar couples
 both to the standard Riemannian volume form given by the
square root of the determinant of the Riemannian metric, as well as to another non-Riemannian volume form given in terms of an auxiliary
maximal rank antisymmetric tensor gauge field. This model provides an exact unified description of both dark energy (via dynamically generated cosmological constant) and dark matter (as a “dust” fluid 
due to a hidden nonlinear Noether symmetry).

In this paper we test the model against Supernovae type Ia experimental data and investigate the future Universe evolution which follows from it. 
Our results show that this model has very interesting features allowing various scenarios  of Universe evolution and in the same time  perfectly fits contemporary observational data. It can describe 
exponentially expanding or finite expanding Universe and moreover, a Universe with phase transition of first kind. The  phase
transition occurs to a new, emerging at some time ground state with lower energy density, which affects significantly the Universe evolution.
\end{abstract}

\newpage

\titlepage

\section{Introduction}
One of the big mysteries of modern cosmology (\cite{cosmo0}) is how to unite the period of inflation after the Big Bang (invoked 
to explain the isotropy of the Cosmic
Microwave  Background  (CMB), the large-scale structures of the Universe, the flatness problem and the horizon problem) and the current period of accelerated expansion confirmed by multiple 
experiments (Planck \cite{cosmo6}, SNe Ia \cite{cosmo1,cosmo2}, the HST key 
project \cite{cosmo3}, the Sloan Digital Sky 
Survey \cite{cosmo5}, WMAP \cite{cosmo4}, large-scale galaxy formations \cite{cosmo7}).

Another problem is to explain the contemporary  matter content of the Universe.
According to the $\Lambda$-CDM model (which can be considered as the Standard Model in cosmology) at the moment the astonishing $95\%$ 
of matter is of unknown type:
it is  dominated by "dark energy"  (or, equivalently --- cosmological constant $\Lambda$) and fewer "cold dark matter". 
Dark energy and cold dark matter are just terms to specify, first, the energy-momentum tensor form of the corresponding matter, second, their scaling properties under Universe expansion and third, the fact that they do not interact electromagnetically, i.e. they are not observable.
The $\Lambda$-CDM model is very successful 
in explaining astronomical observations, but it fails to explain the nature of dark energy and cold dark matter (for example, see \cite{cosmo}). 
While there are numerous 
theories of modified gravity such as the f(R) theory, the Brans-Dicke theory, the dilaton theory etc. each trying to provide a deeper understanding to the dark-content of the Universe, for the moment none of them has proven more successful than the $\Lambda$-CDM model. 

Here  we  work  in  the  framework  of  so  called  "two-measures  theory  of
gravity"  which  can  address  both  of  above  mentioned  problems.   The two-measures theory of gravity has been first proposed in the series of works \cite{ref01}. The essence of the  
 model is that in it the scalar Lagrangian couples both to the standard Riemannian volume-form, as well as to another 
non-Riemannian volume form given in terms of an 
auxiliary maximal-rank antisymmetric tensor gauge field.

Various applications to cosmology of the two-measures theory have been furthermore studied in \cite{15,16_19} and have shown some remarkable results: it generates 
dynamically a cosmological  constant, it gives a natural formulation of a  “quintessential inflation”  scenario in cosmology, it is able to produce two flat regions evolution or a non-singular  
initial “emergent universe” phase preceding the inflationary phase  (giving a new mechanism of dynamical
spontaneous breakdown of supersymmetry in supergravity as well).

Explicitly, in \cite{15} the model was extended to include {\em two} independent non-Riemannian volume forms on the spacetime manifold with the inflation scalar fields potentials of exponential form. 
This choice led to a gravity-matter system invariant under global Weyl-scale symmetry and as a consequence of the equations of motion, to two dimensionful integration constants breaking global 
Weyl-scale invariance. Additionally, the numerical integration of the Friedman equations demonstrated two infinitely large flat regions corresponding to the two different epochs of expansion of the 
Universe -- the early inflation and the current accelerated expansion. 
In \cite{16_19}  an $R^2$ term has been included in the 
model gravitational action, with appropriate field potentials, so that the action is again invariant under global Weyl-scale symmetry and it was shown that one can still obtain the two flat 
regions, but also for some parameter ranges the model produces a non-singular  “emergent universe” solution describing a Universe without a Big Bang. Also, in this case, it was shown that the model 
agrees with the Planck Collaboration data for some choices of the two integration constants. Other interesting developments include \cite{17} where it was applied to produce phase structure 
of 
confinement and deconfinement related to that of MIT bag model of hadrons. In \cite{1603.06231} the model is further extended to include gravity coupled both to the scalar 
inflaton, and also to a SU(2)xU(1) iso-doublet scalar with positive mass squared and without self-interaction, and to SU(2)xU(1) gauge fields. In this case it is shown how the electroweak symmetry 
breaking Higg phase is dynamically generated in the post-inflationary epoch.

In this article, we study numerically the model proposed in \cite{ref2},\cite{GNP}. 
We apply it to the Universe evolution, finding that it can describe (depending on the parameters)  exponentially expanding or finite expanding Universe.
Moreover, we find that the model can describe a Universe with phase transition of first kind.  
Finally we show our
first numerical results for the model fit against Supernovae Ia observational data.

\section{The unified dark energy and dust dark matter model \cite{ref2,GNP}}

In the simplest case of the two-measures theory, one uses the following action $S$:
\begin{align}
  &S=S_{grav}[g_{\mu\nu},\Gamma_{\mu\nu}^{\lambda}]+\int{d^4x(\sqrt{-g}+\Phi(B))L(\phi,X)}\notag\\
  &\Phi(B)=\frac{1}{3\!}\epsilon^{\mu\nu\kappa\lambda}\partial_\mu B_{\nu\kappa\lambda}\notag\\
  &L(\phi,X)=X-V(\phi), \quad X=-\frac{1}{2}g^{\mu\nu}\partial_\mu\phi\partial_\nu \phi.
  \label{TMG}
 \end{align}

Here, the standard Riemannian volume form is $\sqrt{(-g)}$ while the non-Riemannian volume form $\Phi(B)$ is defined by an 
auxiliary antisymmetric gauge field of maximal rank $B_{\nu\kappa\lambda}$.  $L(\phi,X)$ is the general-coordinate invariant Lagrangian of a scalar field $\phi$ with a standard kinetic term $X$, 
which is symmetrically coupled to both measures and $S_{grav}$ is in Palatini formalism.

From the equation of motion for $B$ field we have $L(\phi,X)=-2M$ where $M$ is an integration constant.
The interpretation of $M$ as a dynamically generated cosmological constant can be understood easily from the form of the energy-momentum tensor of the model (\ref{TMG}):
\ba
 T_{\mu\nu}&=&-2Mg_{\mu\nu}+\left(1+\frac{(\Phi(B)}{\sqrt{(-g)}}\right)\partial_\mu \phi \partial_\nu \phi\nn\\
 &=&-2Mg_{\mu\nu}+\rho_0 u_\mu u_\nu,
 \label{tmunu}
\ea
 where $u_\mu=-\frac{\partial_\mu \phi}{\sqrt{2X}}$ and
 $\rho_0=\left(1+\frac{(\Phi(B)}{\sqrt{(-g)}}\right)2X\frac{\partial L}{\partial X}$.
Therefore, we can interpret $T_{\mu\nu}$ as a sum of dark energy and dust contribution with $\rho_{DE}=2M$.

An important feature of the model is that it possesses an additional (hidden) Noether symmetry.

In what follows we consider  a model for which the gravitational part $S_{grav}$ of the action (\ref{TMG}) has the following $f(R)$ form\footnote{In this section, we are using units where the 
Newton constant is $16\pi G_N= 1$}:
 \be
  S_{grav}=\int d^4x\sqrt{-g}(R(g,\Gamma)-\alpha R^2(g,\Gamma))
\ee
It has been proven in \cite{GNP} that this gravitational action is equivalent in the Einstein frame to the following purely kinetic k-essence \cite{k-ess} action with suitably conformally rescaled 
metric $\bar{g}_{\mu\nu}$
\be 
S_k= \int d^4 x \sqrt{-\bar{g}}\left[\bar{R}+\left(\frac{1}{4\alpha}-2M\right)\tilde{X}^2 -\frac{1}{2\alpha}\tilde{X}+\frac{1}{4\alpha}\right]
\ee
where
 $\bar{g}_{\mu\nu}=f_R'\;g_{\mu\nu}$, 
 $f_R'=1-2\alpha R$ and $\tilde{X}=1/f_R'$.
 
\section{The model in Friedman--Lema\^{i}tre--Robertson--Walker metric}
We are working in the so called reduced-circumference polar coordinates, in flat Universe.
In this case the metric can be put in the following form:
\be 
ds^2 = - dt^2 + A(t)^2\left[dr^2 + r^2 \left( d\theta^2 + \sin^2\theta d\varphi^2\right)\right] .
\label{FLRW-metric}
\ee
In the Einstein frame, the Lagrangian is:
\be 
\mathcal{L}=A(t)\left[- \dot{A}(t)^2 + \frac{1}{4} A(t)^2 \left( \frac{1}{\alpha}-\frac{\dot{\phi}^2}{\alpha}+(\frac{1}{4\alpha}-2M) \dot{\phi}^4\right)\right]
\label{lagr}
\ee
The solution of the equation of motion with respect to the field $\phi$ is:
\be 
A(t)^3\left[ -\frac{1}{2 \alpha}\dot{\phi}+(\frac{1}{4\alpha}-2M) \dot{\phi}^3\right]=p_\phi\;\; (=\mathrm{const})
\label{em}
\ee
We obtain the following energy density 
\be 
\rho= \frac{1}{8\alpha}\dot{\phi}^2 +\frac{3}{4}\frac{p_\phi}{A(t)^3}\dot{\phi}-\frac{1}{4\alpha}
\label{ro}
\ee
Let us recall the first Friedman equation
\footnote{In what follows we shall use only the first Friedman equation, because we have Eq.(\ref{em}) which is the equation of state for the model.}
($G_{00}=T_{00}$):
\be \left(\frac{\dot{A}(t)}{A(t)}\right)^2=\frac{1}{6}\rho.
\label{fr}
\ee

We can put Eq.(\ref{em}) in the form
\be 
y^3 +3 \a y + 2 \b = 0
\label{cub}
\ee
where
\ba 
y&=&\dot{\phi}\nn\\
\a&=&-\frac{2}{3 - 24 \alpha M}\nn\\
\b&=& \frac{b}{A(t)^3}\;\;=\;\;-\frac{2\alpha p_\phi}{(1-8\alpha M)A(t)^3}\label{not}.
\ea
Eq.(\ref{cub}) is a cubic equation with real coefficients and therefore, there is always a real solution for every $\a$ and $\b$.
All three solutions of eq.(\ref{cub}) are:
\be
y_i=\left(\sqrt[3]{-1}\right)_i
\frac{\a}{\mathcal{A}}- 
\overline{\left(\sqrt[3]{-1}\right)}_i\mathcal{A}
;\;\; i=1,2,3\ee
where $\mathcal{A}=\sqrt[3]{-\b+\sqrt{\a^3+\b^2}}$ and $\left(\sqrt[3]{-1}\right)_i$ is one of the three roots of $\sqrt[3]{-1}\;\; \left(= -1, \frac{1+i\sqrt{3}}{2}\right.$ and $\left.\frac{1-i\sqrt{3}}{2}\right)$.

The solution $y_1$ is real in the domain 
$\{\a\geq 0\}\cup\{\a < 0 \cap \b<0\}\cup\{\a<0\cap\b>0\cap \a^3+\b^2<0\}$,
$y_2$ is real in the domain 
$\{\a < 0 \cap \b>0\}\cup\{\a<0\cap\b<0\cap \a^3+\b^2<0\}$ and
$y_3$ is real in the domain 
$\{\a<0 \cap\a^3+\b^2<0\}$.
It is easy to see that there is no smooth real solution  in the entire $(\a,\b)$ plane.
Therefore,we may have to jump from one solution of Eq.(\ref{cub}) to another during the Universe evolution.

Here we are interested in of Universe evolution such that at Big Bang ($t=t_{BB}<1$) we have for the metric scaling factor
$A(t_{BB})=0$, and at present epoch ($t=1$) we have $A(1)=1$.
On the $(\a,\b)$ plane the evolution looks like a movement along a half line $\a=\mathrm{const.}$: 
It starts at $(\a,\pm\infty)$ depending on the sign of $b$,
goes through the point $(\a,b)$ and heads to $(\a,0)$.
The solutions of Eq.(\ref{cub}) which are compatible with the early Universe are
$y_1$ for $\{\a \geq 0\}\cup\{\a<0\cap\b<0\}$ and $y_2$ for $\{\a<0\cap\b>0\}$.
So we define a solution $y_b$:
\be 
y_b=\begin{cases}
y_1 \;\;\mathrm{for}\;\; (\a,\b)\in\{\a\geq 0\}\cup\{\a < 0 \cap \b<0\}\\
y_2\;\; \mathrm{for}\;\; (\a,\b)\in \{\a < 0 \cap \b>0\}\label{yp}
\end{cases}
\ee
It is a real solution in the entire $(\a,\b)$ plane which
is smooth except on the half line $(\a<0, 0)$ where it has a jump.
Note however, that this half line is never crossed during the Universe evolution.
This is the solution  we use in our further considerations unless  otherwise explicitly stated.

It is convenient to rewrite the equation of state (\ref{ro})  in terms of notations (\ref{not}):
\be 
\rho = \frac{1}{4|\alpha|}\r= 
\frac{1}{4\alpha}\left(\frac{1}{2} y^2 + \frac{\b}{\a} y -1\right)
\label{ro_p}
\ee
Note that rescaling the time we can ensure
\footnote{$|\alpha|$ eventually goes into Hubble constant.}  
$2|\alpha|/3 =1$,
so the Friedman equation (\ref{fr}) takes the form
\be 
\left(\frac{\dot{A}(t)}{A(t)}\right)^2=\r
\label{freedman}
\ee
Substituting solution (\ref{yp}) into (\ref{ro_p}) we see that not all combinations of constants $\alpha, \a$ and $\b$ are physically acceptable, namely the energy density is non-negative only at 
domain\footnote{There is a sub-region $\a\in(-2/3,0)$ of special interest which will be discussed later.}
$\{\alpha>0\cap  \a<0\}\cup \{\alpha<0\cap \a\geq 0\}$.
So, in what follows, when we discuss $\a<0$ solution we shall understand,  without mentioned it explicitly, that simultaneously $\alpha >0$.
Similarly when we discuss $\a>0$ solution we understand that $\alpha <0$. 

The density $\r$ has a well defined asymptotic value for large A(t)
\ba 
\r & \xrightarrow[A(t)\rightarrow\infty]{}& 1 \;\; \mathrm{for}\;\;\a>0\nn\\
\r & \xrightarrow[A(t)\rightarrow\infty]{}& -\frac{3}{2}\a-1 \;\; \mathrm{for}\;\;\a<0\label{asym}
\ea
which can be interpreted as asymptotic value of the cosmological constant.

\begin{figure}
\centering
\includegraphics[scale=0.3]{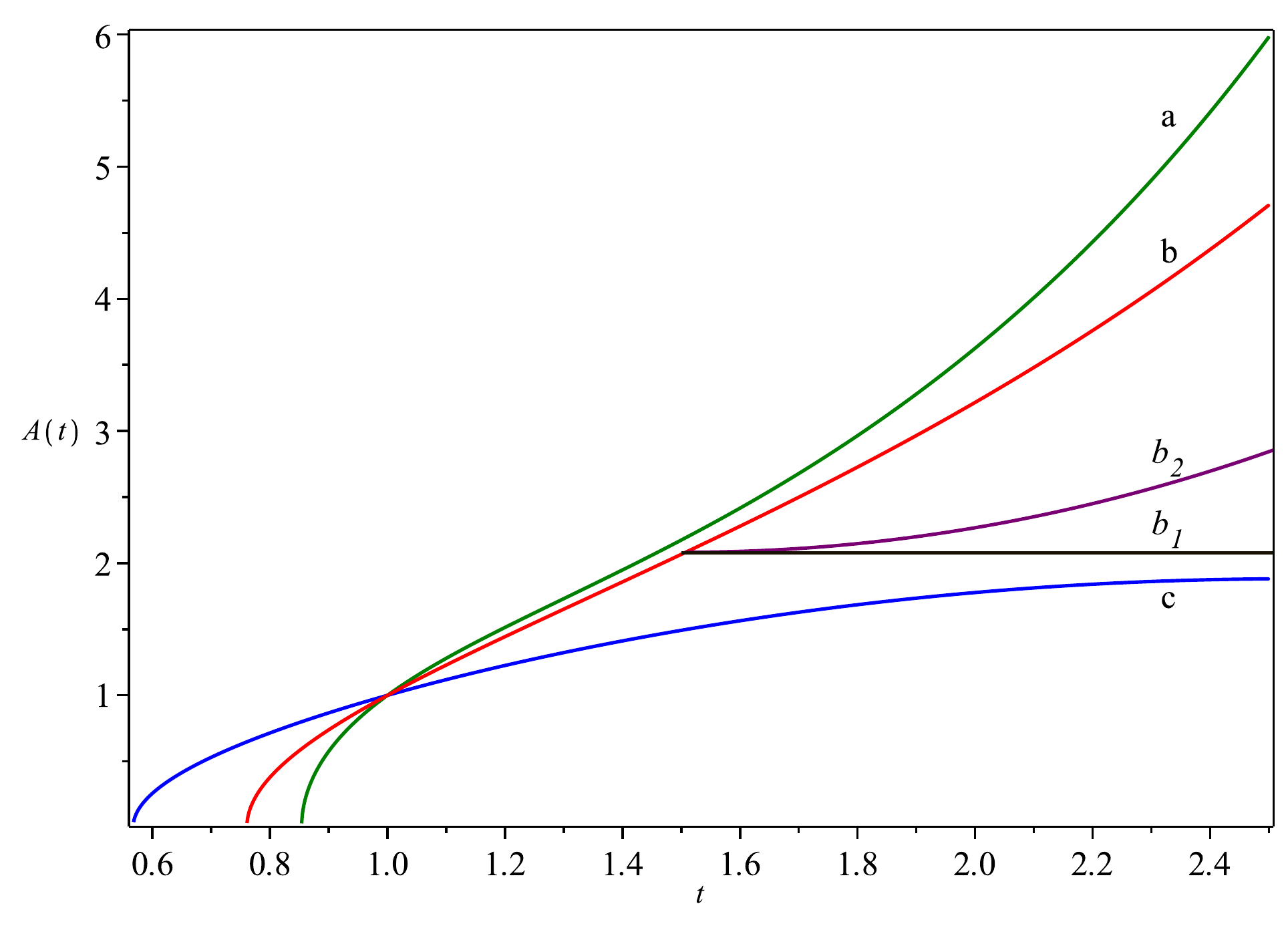}
\caption{Graphics of the $A(t)$ evolution for $\a>0$ ($a$), $\a<-2/3$ ($b$) and $-2/3<\a<0$ ($c$).
Possible evolutions after phase transition: frozen expansion ($b_1$) and re-acceleration ($b_2$).
Note that different values of the parameters change also the moment of the Big Bang, i.e. they 
predict different age of the Universe. \label{f1}}
\end{figure}

We integrate numerically the Friedman Eq.(\ref{freedman}).
The results for the scale factor $A(t)$ evolution are shown on Fig.(\ref{f1}). \footnote{The values of the parameters for each plot are as follows: 
(a) $\a=1, \b=\frac{6}{A(t)^3}$, (b) $\a=-1, \b=-\frac{2}{A(t)^3}$, (c)$\a=-.5, \b=-\frac{0.5}{A(t)^3}$  and ($b_2$): $t_p=1.5074,a_s(t_p)=2.0825$}

Curve ($a$) is typical for $\a>0$.  The initial matter dominated deceleration
and the asymptotic dark energy induced exponential growth $A(t)\sim e^t$ are well visible.

Curve ($b$) is typical for  $\a<0$ case.
Note that according eqs.(\ref{asym}) its final exponential growth can be any (slower or faster than the growth in $\a>0$ case).
Here we show a curve with $\Lambda_{asymptotic} <1$.

A peculiar property of $\a<0$ evolution is that at some moment $t_3$:
\be
t_3:\;\;A(t_3)=\frac{\sqrt[3]{|b|}}{\sqrt[2]{|\a|}}
\ee
we enter a region of $(\a,\b)$ plane where all three  solutions $y_1, y_2$ and $y_3$ are real.
Let us define an additional solution in this region, which we call $y_s$:
\be 
y_s =\begin{cases}
 y_1\;\; \mathrm{for}\;\; b>0\\
 y_2\;\; \mathrm{for}\;\; b<0.\nn
\end{cases}
\ee
Then we can use as independent real solutions $y_b$ (our basic solution), $y_s$  and $y_3$.

It turns out that solution $y_3$ is nonphysical, because the energy density which corresponds to it is always negative.
We denote $\bar{\bar{\rho}}$ the density which corresponds to solution $y_s$. 
Note that $\bar{\bar{\rho}}(t_3)<0$, but because the $\bar{\bar{\rho}}$ asymptotic for large $A(t)$ is the same as that of $\r$ (which is positive) there is a moment $t_p$, such that  
$\bar{\bar{\rho}}(A(t_p))=0$. 
Therefore, for any moment after $t_p$ we have two "states" $\r$ and $\bar{\bar{\rho}}$ of the Universe
\be  
0\leq\bar{\bar{\rho}}<\r\;\; \mathrm{for}\;\; t\geq t_p. 
\ee
This opens the possibility the Universe to undergo "phase transition" or "quenching" to the lower state.
The moment of  the phase transition is crucial for the further evolution: 
If it happens exactly at time $t_p$ the evolution stops ($\bar{\bar{\rho}}=0$) --- this is curve ($b_1$) on Fig.(\ref{f1}).
However, if the jump occurs in any later moment then the evolution will look like at curve ($b_2$). Numerically, this corresponds to using the following boundary conditions: 
$a_b(1)=1, a_s(t_p)=a_b(t_p)$, where the indices $b$ and $s$ refer to the using $y_b$ or $y_s$ in the integration of the Friedman equations.

It has been mentioned earlier that there is  an interesting sub-region  $\a\in(-2/3,0)$.
In this region the early Universe asymptotic value of energy density is positive, but the late Universe asymptotic value is negative.
This has not to be  considered as nonphysical.
Suppose there is a value $A_0$ of the metric scale factor such that $\r(A_0)=0.$
It is easy to see, using Friedman equation (\ref{freedman}), that when $\r$ approaches $0$ from above $A(t)$ tends to a constant: 
\be A(t)\sim \tanh(t)^2 \ee
The Universe never enter the region where $\r$ becomes negative.
As a consequence, the region $\a\in(-2/3,0)$ corresponds to a Universe with finite expansion --- curve ($c$) on Fig.(\ref{f1}).

\section{Supernovae Ia data fit}
Our primary task in this work is to  check the model \cite{GNP} consistency with some of the present observational data.
For this purpose we perform a fit on the data collected by Supernova Cosmology Project \cite{sn}\footnote{Supernovae data are available online at:\\
 http://www.supernova.lbl.gov/Union/figuresSCPUnion2.1\_mu\_vs\_z.txt}.
At the moment the SCP collection contains 580 Supernovae of type Ia
with their distance modulus $\mu$, 
\footnote{The distance modulus $\mu$ is a dimensionless quantity for distance $\mu=5 \log_{10}\left(\frac{d}{10}\right)$ where $d$ is in parsecs.
It is the difference between apparent and absolute star magnitude.}
its experimental uncertainty $\Delta\mu$ and observed red shift $z$.
Surely, we have 
to compare our fit with the existing results and to check our optimization procedure, so we created a reference fit
of the same data using an {\it ad hoc} mixture of dark energy and (dark) matter (Standard Cosmological Model).
Note that we use only basic data provided by SCP, both for $\mu$ and $\Delta\mu$ without taking into account any possible additional systematic errors and without performing subsequent filtering of SN 
Ia which resides away from the fit confidential interval.

There is a relatively simple connection between distance modulus of a star and its red shift (in FLRW metric):
\ba
\mu &=& 5 \log_{10}\left((1+z)\int_0^z dx \frac{A(x)}{A(x)'}\right)=\nonumber\\
&&h+5 \log_{10}\left((1+z)\int_0^z dx \frac{1}{\sqrt{\r(x)}}\right)
\label{dm_b}
\ea
where $h$ is some constant. 
We have used Friedman equation (\ref{freedman}) to obtain the second line of Eq.(\ref{dm_b}) and $A(t)=\frac{1}{1+z} $ to rewrite the energy density in terms of red shift.

The contemporary experimental data quite surely exclude curvature and radiation from matter content of the Universe leaving room only for dark energy and  cold matter.
When we have only these two ingredients the integral in Eq.(\ref{dm_b}) can be take analytically  in terms of Hypergeometric function $_2F_1$.
We use this analytic expression for our reference fit which gives the following result:
\be 
\O_{DE}=0.722,\;\;\O_M=0.278,\;\;\chi^2=562 .\label{st_fit}
\ee
It has to be compared to $\O_{M}=0.277^{+0.022}_{-0.021}$ result of SN fit without systematic listed in Table 7 of \cite{sn}. 
So we find, that our optimization procedure based on the simplex method, is adequate to the considered problem.

\begin{figure}[htb]
\centering
\includegraphics[scale=0.5]{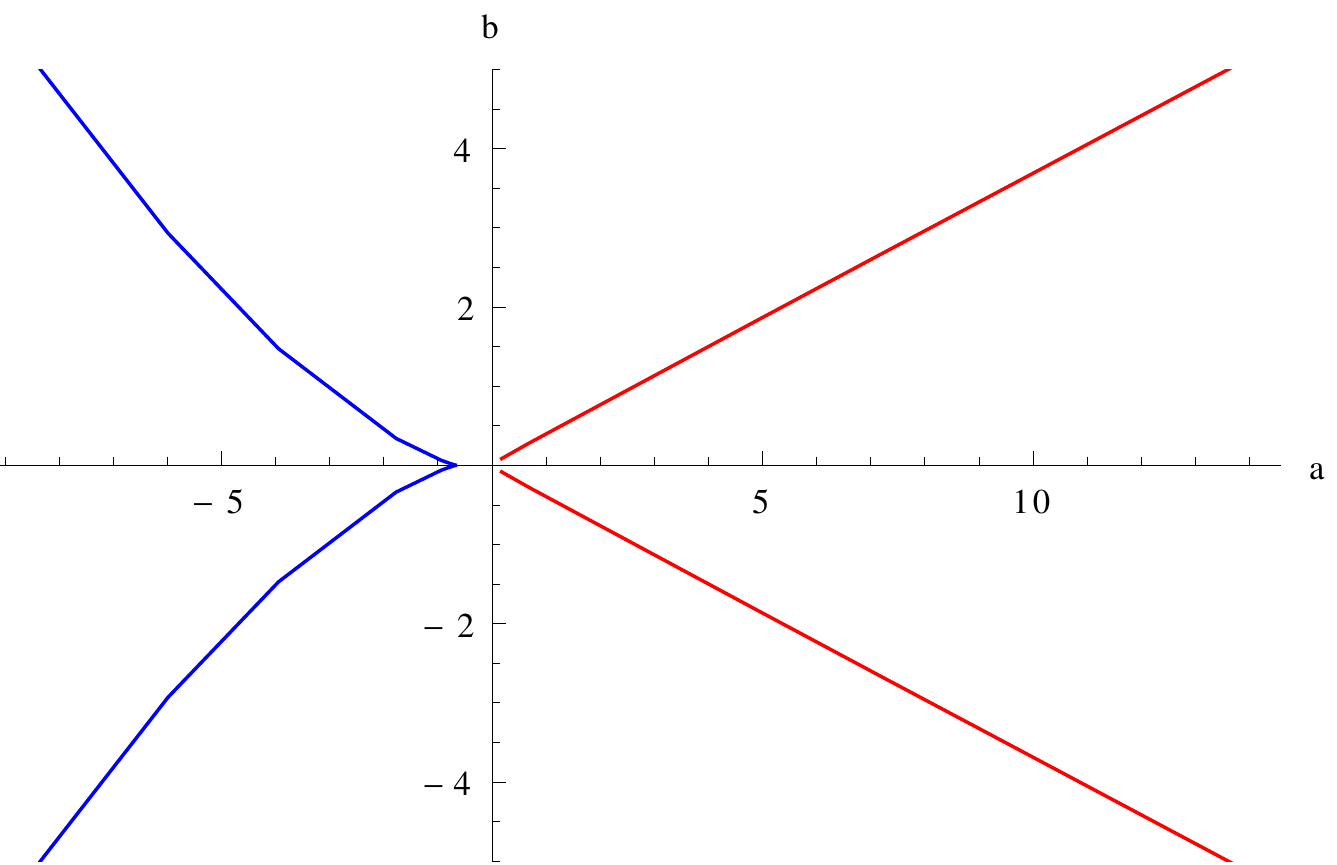}
\caption{\label{f2} Best fit families on the parametric plane.}
\end{figure}

We use numerical integration of the integral in Eq.(\ref{dm_b}) when we test the model \cite{GNP}.
It turns out that the best fit of Supernovae data for the proposed model is not unique.
We observe two one-parametric families of best fits, shown on Fig.(\ref{f2}) as two curves on the  model parameters plane.
Any point $(\a,b)$ on each curve  gives  approximately the same $\chi^2$ and  indistinguishable  $\mu(z)$ function.
Our result is:
\ba 
\chi^2 &\sim& 562 \;\;\mathrm{for} \;\; \a<-2/3, \nn\\
\chi^2 &\sim& 578 \;\;\mathrm{for} \;\; \a>1. \label{o_fit}
\ea
We observe a small variation of the $\chi^2$ in the interval $\a\in(0,1)$ where  $\chi^2$ slightly decreases when $\a\rightarrow 0$  ($\chi^2\sim 568$ for $\a=.18$), but this could be just a numerical 
artifact.
Both families have $\chi^2$ very close to that of Standard model (Eq.(\ref{st_fit})), even the one  with $\a <-2/3$ has $\chi^2$ with a fraction of unit better.
Note that there is no good fit for $\a\in(-2/3,0)$ (finite expansion Universe).

The question what sands behind the observed families of best fit solutions has not clear answer.
For the family $\a<-2/3$ the constant $h$ in Eq.(\ref{dm_b}) also depends on the family parameter (we can use $b$ as such), so the existence of the  family may reflect the fact that we have not fixed 
the time scale.
However, this is not the case with the $\a>0$ family, where $h$ practically does not depend on $b$.

\begin{figure}[htb]
\centering
\includegraphics[scale=.5]{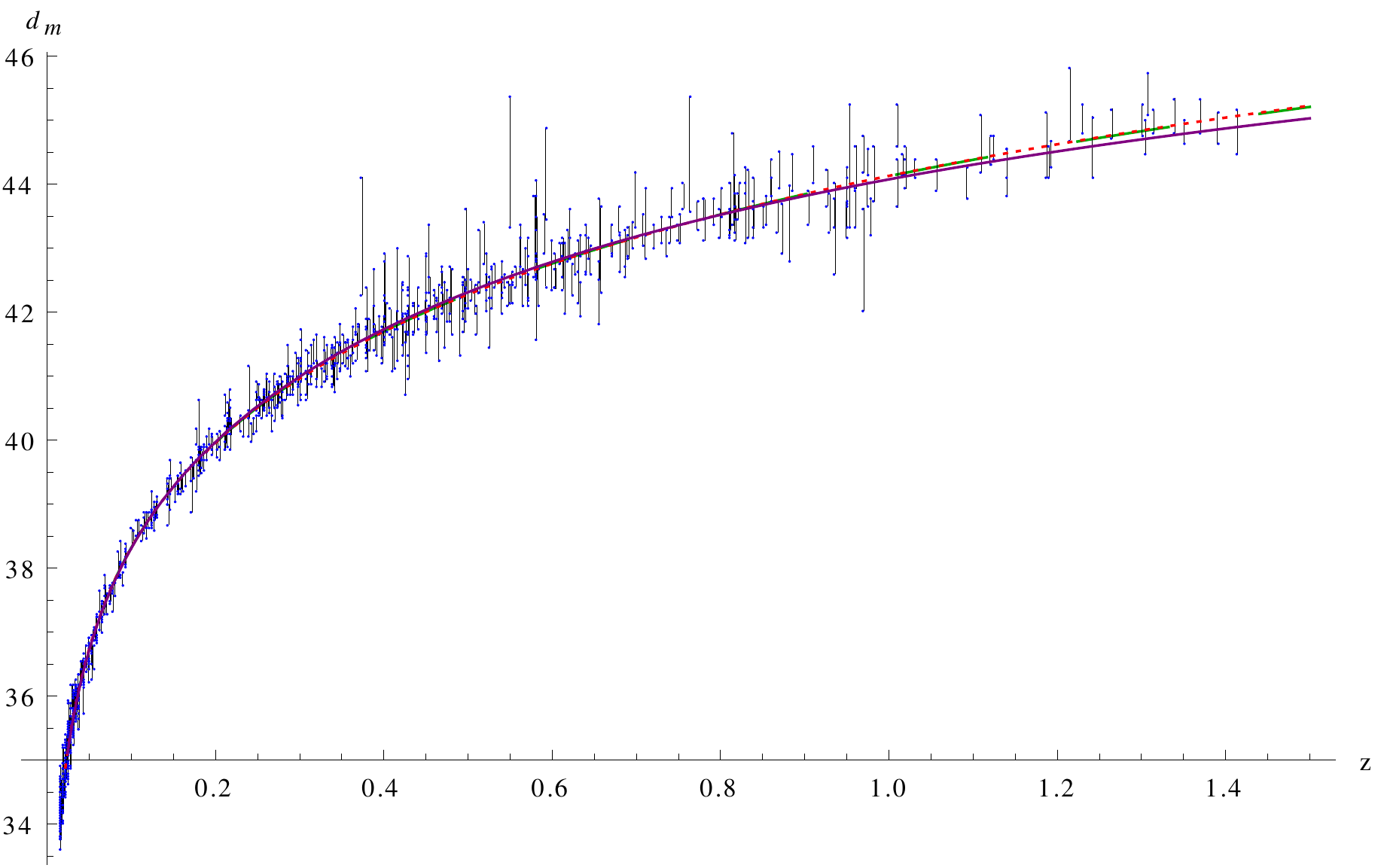}
\caption{\label{f3} Supernovae data against Standard model fit (dotted line), $\a<-2/3$ fit (dashed line) and $\a>1$ fit (solid line). }
\end{figure}

Our best fit $\mu(z)$ functions are shown on Fig.(\ref{f3}) against Supernovae data and  standard fit (\ref{st_fit})).
All curves are very close to each other and for large interval of $z$ overlap.

\section{Conclusions}

In this article, we have examined the applications of the k-essence theory to cosmology. By studying the domains of validity of the theory parameters $\left[\a,b\right]$ and the energy density dependency on them, we have been able to obtain both Universes with and without phase transition. We have used the latest SN Ia data to  fit the model.
We have obtained two one-parametric families of solutions, 
each fitting the observations, showing that the choice of values for $\left[\a,b\right]$ is not unique. 
The obtained fits are with $\chi^2$ same as that given by $\Lambda$-CDM model while at the same time we have huge parameter flexibility allowing additional model fine tuning.
The model allows a phase transition from evolution which is compatible with SN Ia data  to occur after our epoch, thus the scenarios for the further evolution of our Universe are still open.
An analogous phase transition mechanism can be used to explain the transition from inflation to matter dominated epoch of the Universe.

\section*{Acknowledgments}

It is a pleasure to thank E. Nissimov and S. Pacheva for the discussions. 

The work is supported by BAS contract DFNP -- 49/21.04.2016.
 MS also acknowledge the support of BNSF under grant DFNI-T 02/6.

\end{document}